\documentclass[prl,aps,twocolumn,showpacs,superscriptaddress]{revtex4}
\usepackage{graphicx,amsmath,amsfonts,latexsym,bm}

  \def\cH{{\cal H}}    
 
 \def\br{{\bf r}}

\begin{document} 
\title{Transition from a one-dimensional to a
  quasi-one-dimensional state in interacting quantum wires}

\author{Julia S. Meyer}

\affiliation{Department of Physics, The Ohio State University, Columbus,
  Ohio 43210, USA}

\author{K. A. Matveev}

\affiliation{Materials Science Division, Argonne National Laboratory,
  Argonne, Illinois 60439, USA}

\author{A. I.  Larkin}
 
\affiliation{W.I. Fine Theoretical Physics Institute, University of
  Minnesota, Minneapolis, Minnesota 55455, USA}

\date{\today}
 
 
\begin{abstract} 
  Upon increasing the electron density in a quantum wire, the
  one-dimensional electron system undergoes a transition to a
  \textit{quasi}-one-dimensional state.  In the absence of interactions
  between electrons, this corresponds to filling up the second subband of
  transverse quantization, and there are two gapless excitation modes
  above the transition.  On the other hand, strongly interacting
  one-dimensional electrons form a Wigner crystal, and the transition
  corresponds to it splitting into two chains (zigzag crystal).  The two
  chains are locked, so their relative motion is gapped, and only one
  gapless mode remains.  We study the evolution of the system as the
  interaction strength changes, and show that only one gapless mode exists
  near the transition at any interaction strength.
\end{abstract} 
 
\maketitle
 
Transport properties of quantum wires have attracted much attention over
recent years~\cite{Wees, Thomas, Reilly, Cronenwett, DePicciotto,
  Rokhinson, Auslaender}.  Due to their quasi-one-dimensional structure,
conductance is expected to be quantized in units of the conductance
quantum $G_0=2e^2/h$, where $e$ is the elementary charge and the factor of
2 accounts for spin degeneracy.  This property of non-interacting
electrons is insensitive to the inclusion of interactions within the
Luttinger liquid description.  However, a number of experiments show
deviations from perfect conductance quantization, such as the so-called
0.7-structure observed below the first plateau \cite{Thomas, Reilly,
  Cronenwett, DePicciotto, Rokhinson}.  These experiments have stimulated
much theoretical interest in the physics of one-dimensional conductors not
captured by the Luttinger-liquid theory, such as that of the
spin-incoherent regime characterized by very weak coupling of the electron
spins \cite{Matveev,Cheianov,Fiete}.  
Here we consider another important problem in this category: the
transition of the one-dimensional electron system in a quantum wire into a
quasi-one-dimensional state.

Whether or not an electron system can be viewed as one-dimensional
crucially depends on the strength of interaction~\cite{us}.  In the
absence of interactions, electrons occupy subbands of transverse
quantization, and the system is one-dimensional until the chemical
potential reaches the bottom of the second subband.  On the other hand, at
strong interactions, the electrons form a Wigner crystal, and the subband
picture is no longer applicable.  The system remains one-dimensional until
the interaction energy overcomes the confining potential, and the crystal
splits into two chains, forming a zigzag structure~\cite{Piacente}.

\begin{figure}[b]
 \resizebox{.45\textwidth}{!}{\includegraphics{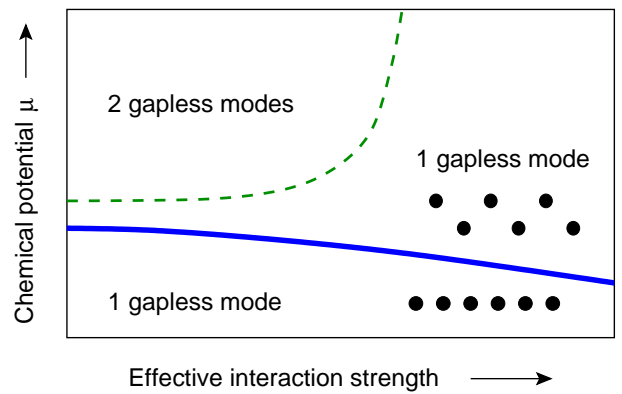}}
\caption{\label{fig-phase} Phase diagram of the electronic system in a
  quantum wire. The interaction strength can be tuned by varying the
  confining potential.  In the vicinity of the transition to a
  quasi-one-dimensional state (the lower line) the system supports only
  one gapless excitation mode at any interaction strength.  The state with
  two gapless excitation modes is expected only at relatively weak
  interactions.}
\end{figure}

There is an obvious difference in the behavior of the system in the
vicinity of the transition between the limiting cases of the Wigner
crystal at strong interactions on the one hand and non-interacting
electrons on the other.  In a zigzag Wigner crystal, the two chains are
locked, and only one gapless mode (the plasmon) remains.  By contrast, in
the case of non-interacting electrons the two subbands are independent and
therefore represent two gapless modes.  In this paper, we address the fate
of the gapped mode in the vicinity of the transition as the interaction
strength varies. In particular, we show that a gap exists at any
interaction strength, Fig.~\ref{fig-phase}.  However, the nature of the
transition changes as the interaction strength is varied.

For simplicity we consider spinless electrons and assume that they
interact via long-range Coulomb repulsion,
\begin{equation}
V_{\rm int}=\frac{e^2}{2\epsilon }\sum_{k\neq l}\frac 1{|\br_k-\br_l|}.
\end{equation}
Here $\br_k$ are the two-dimensional position vectors of the
electrons, and $\epsilon$ is the dielectric constant of the material.

If the electrons in the wire are confined to one dimension by a strong
external potential $V_{\rm conf}(y_k)$, their physics is controlled by the
one-dimensional electron density $n_e$.  Since at $n_e\to0$ the kinetic
energy of an electron $\sim (\hbar^2/m)n_e^2$ scales to zero faster than
the interaction energy $\sim (e^2/\epsilon)n_e$, at low densities the
Coulomb repulsion dominates.  In this limit electrons behave classically.
In order to minimize their mutual interaction, they form a periodic
one-dimensional structure---the so-called Wigner crystal.  At small but
finite density, quantum fluctuations smear the long-range order
\cite{schulz}, but the short-range order remains as long as the distance
between electrons $n_e^{-1}$ is greater than the Bohr radius
$a_B=\epsilon\hbar^2/me^2$.

The above picture is valid if the width $w$ of the wire is small, $w\ll
a_B$.  In wider wires, or, equivalently, at stronger interactions, the
opposite regime $w\gg a_B$ can be achieved.  In this case the electrons
may form a two-dimensional structure while remaining essentially
classical.  The structure of the Wigner crystal in this regime can be
studied in detail (cf.\ Ref.~\onlinecite{Piacente}), if the confining
potential is quadratic,
\begin{equation}
V_{\rm conf}=\frac12m\Omega^2\sum_k y_k^2.
\label{eq:V_conf}
\end{equation}
Here the frequency $\Omega$ determines the width of the wire,
$w\sim\sqrt{\hbar/m\Omega}$.  The positions of all electrons are found by
minimizing the energy $V_{\rm int}+V_{\rm conf}$ over $\br_k$ keeping the
one-dimensional electron density $n_e$ fixed.  The geometry of the
classical crystal is controlled by the dimensionless electron density $\nu
=n_e r_0$, where $r_0=(2e^2/\epsilon m\Omega^2)^{1/3}$ is the sole length
scale of the problem \cite{Piacente}.  At $\nu$ below the critical value
$\nu_c=\sqrt[3]{4/7\zeta(3)}\approx 0.780$ the energy is minimized when
electrons form a one-dimensional crystal, in which $x_k=k/n_e$ and
$y_k=0$.  At $\nu>\nu_c$ the crystal splits into two rows.  The distance
between rows vanishes at the transition; just above the critical value,
when $\delta\nu=\nu-\nu_c\ll1$, it grows as
$c=(\sqrt{{24}/{93\zeta(5)}}/\nu_c^{2})\sqrt{\delta \nu}$ (in units of
$r_0$).

Let us consider the low-energy phonons in the zigzag Wigner crystal.
Regardless of the density, the crystal has the usual plasmon excitation
with acoustic spectrum.  In the limit of zero wavevector, this excitation
corresponds to translation of the crystal along the wire, $\delta
x_k=\eta$, $\delta y_k = 0$ for any $k$.  In addition, at the zigzag
transition point a transverse soft mode appears, for which $\delta x_k=0$
and $\delta y_k=(-1)^k \varphi$.  One can easily show that near the
zigzag transition, when $\delta\nu\ll1$, the coupling of the two
low-energy excitation modes is weak, and they can be treated separately.
The action describing the soft transverse mode takes the form
\begin{equation}
  S=A\hbar\sqrt{\frac{r_0}{a_B}}\int d\tau\,dx\left[(\partial_\tau\varphi)^2+
    (\partial_x\varphi)^2-\delta\nu\,\varphi^2+\varphi^4\right].
  \label{eq:action}
\end{equation}
Here $x$, $\tau$, and the field $\varphi$ have been rescaled so as to
yield the simplest form of the action possible.  This form, as well as our
results from this point on, are not sensitive to the exact shape of the
confining potential.  In case of the parabolic potential (\ref{eq:V_conf})
the constant $A=[7\zeta(3)]^{3/2}\sqrt{\ln2}/31\zeta(5)$.

Above the classical transition point, i.e., at $\delta\nu>0$, the
transverse mode becomes unstable.  This corresponds to the formation of
the zigzag structure.  To stabilize the system we keep the quartic term
$\varphi^4$.  Quantum fluctuations affect both the position and the nature
of the phase transition.  To determine their effect, it is helpful to
fermionize the problem.  To this end we rediscretize the coordinate $x$
and consider a set of particles moving in a double-well potential
$-\lambda \varphi_j^2 +\varphi_j^4$ with nearest neighbor interaction
$(\varphi_j-\varphi_{j+1})^2$ between them.  At sufficiently large
$\lambda$ each particle is almost completely localized in one of the
minima, and its position can be described by a spin operator, $\varphi_j =
\pm\sqrt{\lambda/2}= \sqrt{\lambda/2}\,\sigma_j^z$.  In terms of these
pseudo-spin variables the Hamiltonian contains two terms:
$-t\sum_j\sigma^x_j$ describing tunneling between the two minima of the
double-well potential and $-v\sum_j\sigma^z_j\sigma^z_{j+1}$ accounting
for the nearest-neighbor interactions.  Rotating $\sigma^x\to-\sigma^z$
and applying the Jordan-Wigner transformation, one obtains the Hamiltonian
\begin{eqnarray}
\cH_{\rm f}=\sum_j\Big[2t\,a_j^\dagger
  a_j-v\Big(a_j^\dagger-a_j\Big)\Big(a_{j+1}^\dagger+a_{j+1}\Big)\Big].
\label{eq-JW}
\end{eqnarray}

The above procedure has enabled us to convert the bosonic problem
(\ref{eq:action}) to that of non-interacting spinless fermions
(\ref{eq-JW}).  Since the number of fermions is not conserved, the
Hamiltonian should be diagonalized by performing a Bogoliubov
transformation.  As a result one easily finds that the excitation spectrum
of the Hamiltonian (\ref{eq-JW}) has a gap $\Delta$ that vanishes when
$t=v$.  We identify this point with the phase transition from the
one-dimensional state of the wire at $t>v$ when all the fermionic states
in the Hamiltonian (\ref{eq-JW}) are empty, to the quasi-one-dimensional
state at $v>t$, in which fermionic states describing the transverse
degrees of freedom in the wire are filled, but possess a spectral gap.
Near the transition the gap behaves linearly, $\Delta=2|v-t|$.

In experiments with quantum wires, the transition from a one-dimensional
to a quasi-one-dimensional state is observed when the chemical potential
$\mu$ of electrons is tuned by applying a voltage to the gate controlling
the electron density.  The parameters $t$ and $v$ of the Hamiltonian
(\ref{eq-JW}) are expected to be non-singular functions of $\mu$.  The
transition occurs at the critical value $\mu_c$, defined as a solution of
the equation $t(\mu)=v(\mu)$.  The gap in the excitation spectrum is then
linear in the distance from the transition,
\begin{equation}
  \label{eq:Ising_gap}
  \Delta\propto |\mu - \mu_c|.
\end{equation}
To better understand the nature of this transition, it is helpful to
consider the well-known mapping between phase transitions in
$d$-dimensional quantum systems and $(d+1)$-dimensional classical models
\cite{Vaks-Larkin}.  In particular, the phase transition in the
one-dimensional quantum model (\ref{eq:action}) is equivalent to that in
the two-dimensional classical Ising model \cite{Polyakov}.  In this
mapping the gap $\Delta$ becomes the inverse correlation length $r_c^{-1}$
of the Ising model, and the scaling $r_c\propto|T-T_c|^{-1}$, well-known
from the exact solutions, is equivalent to Eq. (\ref{eq:Ising_gap}).  The
relation between these phase transitions can be made more explicit by
noticing that our Hamiltonian (\ref{eq-JW}) essentially coincides with the
transfer matrix \cite{Mattis} of the Ising model near the transition
point.

\begin{figure}[t]
 \resizebox{.45\textwidth}{!}{\includegraphics{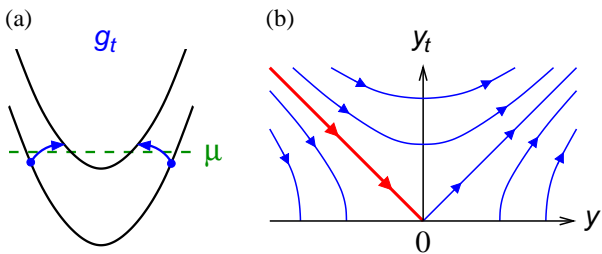}}
\caption{\label{fig-int} (a) The scattering processes transferring
  pairs of particles between the two subbands.  (b) The flow diagram for
  the renormalization group (\ref{eq:RG}).}
\end{figure}

In the discussion leading to the result (\ref{eq:Ising_gap}), the
interactions in the quantum wire were assumed to be very strong.  To
explore the fate of the gap as the interaction strength is reduced, we now
turn to the case of weak interactions.  In this case the transition to the
quasi-one-dimensional state occurs when electrons start populating the
second subband of transverse quantization in the wire.  In this regime one
can neglect the presence of the other subbands and present electron
wavefunctions as $ \psi(x,y)= \psi_1(x)\chi_1(y) +\psi_2(x)\chi_2(y), $
where $\chi_{1,2}(y)$ are the first and second eigenstates in the
confining potential.  Weak interactions between electrons lead to coupling
of the two subbands.  The low energy properties of the system are
described by four interaction constants.  The first three constants,
$g_1$, $g_2$, and $g_x$, correspond to density-density interactions in the
first subband, the second one, and between the two subbands, respectively.
The fourth coupling constant $g_t$ accounts for the possibility of
transfer of pairs of electrons from one subband to the other,
Fig.~\ref{fig-int}(a).

It is well known that in one-dimensional systems density-density
couplings renormalize the velocities of the acoustic low-energy excitations,
but do not lead to the emergence of spectral gaps \cite{Giamarchi}.  On the
other hand, the coupling $g_t$ creates and destroys pairs of electrons in
each subband and thus can, in principle, lead to a BCS-like gap in the
spectrum.  Since the coupling constants in weakly interacting
one-dimensional electron systems acquire logarithmic renormalizations at
low energies \cite{Giamarchi}, the existence of a gap is determined by
the scaling of $g_t$.

The renormalization group equations for the four coupling constants have
been derived in Ref.~\onlinecite{Ledermann}.  As the bandwidth of the
problem is scaled from $D_0$ down to $D$, the renormalization of the
coupling constants can be found by solving the system of two coupled
equations
\begin{equation}
y_t'= yy_t,
\quad
y'= y_t^2.
\label{eq:RG}
\end{equation}
Here the derivatives are with respect to $\xi=\ln(D_0/D)$,
\begin{eqnarray*}
  \label{eq:y_t}
  y_t &=& \frac{g_t}{\pi\hbar} 
        \sqrt{\frac{(v_{F1}+v_{F2})^2+4v_{F1}v_{F2}}
                   {2v_{F1}v_{F2}(v_{F1}+v_{F2})^2}},
\\
  y&=&-\frac1{2\pi\hbar } \left(\frac{g_1}{v_{F1}} + \frac{g_2}{v_{F2}} -
  \frac{4g_x}{v_{F1}+v_{F2}}\right),
\end{eqnarray*}
$v_{F1}$ and $v_{F2}$ are the Fermi velocities in the two subbands.

The renormalization group flow corresponding to the equations
(\ref{eq:RG}) is shown in Fig.~\ref{fig-int}(b).  To find the initial
values of $y_t$ and $y$ we compute the coupling constants in first order
in the interaction strength.  Assuming that the Coulomb interactions
between electrons are screened by a gate at distance $d$, in the limit of
low electron density in the second subband we find with logarithmic
accuracy $g_1^{(0)}=g_x^{(0)}= 2e^2\ln(k_{F1}d)$ and $g_2^{(0)}\sim
e^2(k_{F2}d)^2 \ln(1/k_{F2}d)$.  It is important to note that
$g_2^{(0)}/v_{F2}$ vanishes when approaching the transition.  This is a
consequence of the Pauli principle.  When the average distance between
electrons is large, interactions between them are effectively local.
Then, as identical fermions never occupy the same place, electrons
essentially do not interact.  From these estimates we conclude that
$y^{(0)}$ is positive, $y^{(0)}\simeq (3e^2/\pi\hbar
v_{F1})\ln(k_{F1}d)$, and according to the flow diagram
Fig.~\ref{fig-int}(b), the interaction constant $g_t$ scales to infinity.
Consequently, the system develops a spectral gap.  To find its value, we
estimate $g_t^{(0)}$ near the transition as $g_t^{(0)}\sim
e^2k_{F2}/k_{F1}$ and obtain
\begin{equation}
  \label{eq:weak_gap}
  \Delta\propto(\mu-\mu_c)^{\alpha}.
\end{equation}
Thus the gap in the spectrum of transverse excitations of the wire exists
not only when the interactions are strong, but also when they are weak.
Unlike the case of strong interactions (\ref{eq:Ising_gap}), at weak
coupling the power-law dependence (\ref{eq:weak_gap}) has a very large
exponent $\alpha=(4y^{(0)})^{-1}\gg1$.

To gain insight into the evolution of the transition between the two
limiting cases, we derive the effective Hamiltonian of the system at
intermediate interactions.  Since the interactions between electrons in
the lower subband are no longer weak and only their properties near the
Fermi level are important, it is convenient to use the bosonization
approach \cite{Giamarchi}.  On the other hand, as we discussed, near the
transition interactions between electrons in the second subband are
negligible, $g_2\to0$.  Furthermore, the curvature of their spectrum is
important in this regime, so the description in terms of fermionic
operators is more appropriate.  The non-vanishing density-density
interactions can still be described by the constants $g_1$ and $g_x$,
although their values may no longer be computed in first-order
perturbation theory.  Under these conditions the Hamiltonian has the form
\begin{eqnarray}
\cH&=&\frac{\hbar v_{F1}}{2\pi}\int dx
    \left((\partial\theta)^2+\frac{(\partial\phi)^2}{K^2}\right)
-\frac{\hbar^2}{2m}\int dx\,\psi^\dagger
      \partial^2\psi
\nonumber\\
   &&+\gamma_t
      \int dx\left\{[(\partial\psi)\psi-\psi\partial\psi]
                    e^{2i\kappa\theta(x)}
                    +{\rm h.c.}\right\}. 
\label{eq:boson-fermion}
\end{eqnarray}
Here the bosonic fields $\phi(x)$ and $\theta(x)$ describe the density
excitations in the first subband, $K=(1+g_1/\pi\hbar v_{F1})^{-1/2}$ is
the respective Luttinger liquid parameter, $\psi$ is the electron
destruction operator for the second subband, the constant $\gamma_t\sim e^2$,
and $\kappa=1-K^2g_{x}/\pi\hbar v_{F1}$.  In deriving
Eq.~(\ref{eq:boson-fermion}) we performed a unitary transformation
\cite{balents} which removed the density-density coupling between the two
subbands and changed the phase factor $e^{2i\theta}$ in the last term to
$e^{2i\kappa\theta}$.

The Hamiltonian (\ref{eq:boson-fermion}) interpolates between the limits
of weak and strong interactions.  In the weak coupling limit $\kappa\to1$,
and a simple scaling analysis recovers the renormalization group equations
(\ref{eq:RG}).  At strong interactions the problem of evaluating the
coupling constants $g_1$ and $g_x$ is non-trivial, but in the regime
$e^2/\hbar v_{F1}\ll 1 \ll (e^2/\hbar v_{F1}) \ln(k_{F1}d)$ one can still
use our earlier estimates $g_1=g_x=2e^2\ln(k_{F1}d)$ and conclude that
$\kappa\to0$.  Interestingly, in this case the bosonic and fermionic parts
of the Hamiltonian (\ref{eq:boson-fermion}) decouple, with the latter
becoming equivalent to Eq.~(\ref{eq-JW}).

One can use the Hamiltonian (\ref{eq:boson-fermion}) to discuss the
evolution of the gap with varying interaction strength.  In the limit of
strong interactions, when $\kappa=0$, the magnitude of the pairing term
scales to zero near the transition as $k_{F2}$, i.e., slower than the
Fermi energy $E_{F2}\propto k_{F2}^2$. In this regime, the gap equals the
Fermi energy, $\Delta=E_{F2}$, cf.\ Eq.~(\ref{eq:Ising_gap}).  At strong
but finite interactions the pairing term suffers additional power-law
suppression at $k_{F2}\to0$ because of the factor $e^{2i\kappa\theta}$.
However, as long as it scales slower than the Fermi energy, the magnitude
of the gap remains $\Delta=E_{F2}$.  At weaker interactions, when $\kappa$
exceeds a certain critical value, the pairing term scales to zero faster
than the Fermi energy.  In this regime the gap develops in a small
vicinity of the Fermi points, and its dependence on chemical potential is
given by a non-universal power law (\ref{eq:weak_gap}) with exponent
$\alpha>1$.  A more detailed theory of the transition at intermediate
interaction strengths will be reported elsewhere \cite{unpub}.

Our results are summarized in the phase diagram Fig.~\ref{fig-phase}.  The
electron system in a quantum wire remains one-dimensional and has a single
acoustic excitation branch until the chemical potential reaches a certain
critical value $\mu_c$. At the critical point there is a second gapless
mode, and the system can no longer be viewed as one-dimensional.  At
$\mu>\mu_c$ the second mode develops a gap $\Delta\propto
(\mu-\mu_c)^\alpha$ with exponent $\alpha=1$ at strong interactions, but
very large $\alpha$ at weak coupling.  At weak interactions, as the
chemical potential is increased further, the residual interactions $g_2$
grow, and the gap disappears.  This happens when the electron density in
the second subband is still very small, of order $(k_Fd^2)^{-1}$, long
before the third transverse mode appears.  We see no physical reason for
the gap to disappear at strong coupling.  In experiments, the presence of
a gap will affect the temperature dependence of the conductance which is
expected to show activated behavior even above the transition into the
quasi-one-dimensional state.  The doubling of the zero-temperature
conductance (from $e^2/h$ to $2e^2/h$ for spinless electrons) occurs at
the upper line in Fig.~\ref{fig-phase}.

\begin{acknowledgments} 
  We would like to acknowledge helpful discussions with A. Furusaki, T.
  Giamarchi, L.~I. Glazman, A.~D.  Klironomos, K. Le Hur, and O.~A.
  Starykh.  This work was supported by the U.S. Department of Energy,
  Office of Science, under Contract No.~W-31-109-ENG-38.  We thank the
  Aspen Center for Physics, where part of this work was done, for
  hospitality.
\end{acknowledgments}


\begin{thebibliography}{}
\bibitem{footnote} This result differs from the one obtained in
  Ref.~\onlinecite{Ledermann}, where no gap was found at $v_{F2}\to0$
  because of the assumption $g_t^{(0)}=g_1^{(0)}=g_2^{(0)}=g_x^{(0)}$.
  This choice of the initial conditions is not appropriate in our case.
\end{thebibliography}
\end{document}